\documentclass[preprint,showpacs,preprintnumbers,amsmath,amssymb,floatfix]{revtex4}

\usepackage{graphicx}
\usepackage{dcolumn}
\usepackage{bm}

\newcommand{\Vr}{{\bm r}}
\newcommand{\Vp}{{\bm p}}
\newcommand{\Vk}{{\bm k}}

\newcommand{\Vn}{{\bm n}}

\begin{document}

\title{Effective Field
Theory of Nucleon-Nucleon Scattering \\
on Large Discrete Lattices}

\author{Ryoichi Seki${}^{1,2}$ and U. van Kolck${}^3$}

\affiliation{ ${}^1$ Department of Physics and Astronomy,
California State University, Northridge, Northridge, CA 91330 \\
${}^2$ W.K. Kellogg Radiation Laboratory, California Institute of
Technology, Pasadena, CA 91125\\
${}^3$ Department of Physics, University of Arizona, Tucson, AZ
85721}

\date{\today}

\begin{abstract}
Nuclear effective field theory is applied to the effective range
expansion of S-wave nucleon-nucleon scattering on a discrete lattice.
Lattice regularization is demonstrated to yield the effective range
expansion in the same way as in the usual continuous open space.
The relation between the effective range parameters and the potential
parameters is presented in the limit of a large lattice.
\end{abstract}


\maketitle

\section{Introduction}

In the last several years nuclear effective field theory (EFT) has been
applied extensively to low-energy nucleon-nucleon interactions and
to few-nucleon systems \cite{review}.
At energies below 1 GeV or so, quantum
chromodynamics (QCD) reduces to a hadronic theory
containing all interactions allowed by the symmetries of the theory.
At the very lowest energies,
the interactions are of contact type among nucleons,
with arbitrary number of derivatives.
For the two-nucleon system,
the nuclear EFT has
established a concrete systematic foundation for the traditional
description represented by the effective range expansion (ERE), even
with large S-wave scattering lengths generated by bound or
nearly bound states \cite{vankolck,ksw,gegelia}.
In systems with more than two nucleons
few-body contact forces are present
---a three-body force appears already in leading order---
and the EFT provides a well-defined, successful extension of the
ERE \cite{three,four}. At higher energies, pions need to be
accounted for explicitly in the theory. In this case, the EFT goes
beyond the ERE even in the two-nucleon system, albeit at the cost
of a much more complicated renormalization structure
\cite{review}.

Once the leading few-body interactions are determined from
few-body systems, the main goal of the EFT program is to predict
the structure of larger nuclei. Before tackling heavy nuclei, one
would like to be able to predict the properties of infinite
nuclear matter. This requires a method of solution whose errors
are not larger than EFT truncation errors. A few years ago it was
suggested that this could be achieved by putting nucleons on a
spatial lattice and using Monte Carlo methods to compute the
partition function \cite{mksv}. In this first, exploratory
investigation we considered two-nucleon contact interactions only,
with parameters adjusted to nuclear matter properties.
Subsequently, works have appeared that extend this approach in
various directions
\cite{chen,shailesh,ASK,deanlee1,deanlee2,deanlee3,aurel}. As yet,
however, a full application of EFT to nuclear matter has not been
carried out. In this work we take the first step toward this goal
by examining the effective range expansion of nucleon-nucleon
scattering on a discrete lattice. The extension of this work to
the determination of thermal properties of neutron
matter from parameters from few-body physics is
currently underway \cite{AS}.

As in any field theory, the parameters that appear in the EFT
Lagrangian or Hamiltonian are not directly observable,
since the separation between them and the high-momentum components
of loops is arbitrary. This separation is the regularization
procedure, such as momentum cutoff and dimensional.
Because the separation is arbitrary, relations
among observables should not
depend on the regularization scheme.
The relation between EFT parameters and observables, on the other hand,
does depend on the regularization,
and some regularization schemes are more convenient to apply
than others.
The program of predicting many-body properties from few-body physics
requires that the relation between parameters and observables
be known within the regularization
scheme employed in the solution of the many-body problem.
Placing nucleons in a lattice is a choice of a regularization
scheme.

In this work, we examine lattice regularization for the
effective field theory
on a discrete three-dimensional cubic lattice of a large size.
The special aspect associated with the use of a
lattice is that the nucleons are interacting in a closed space,
different from scattering of nucleons in the open space. On
this issue, the method of L\"{u}scher
\cite{luescher} is well known in lattice QCD,
and it has been also studied for the
nucleon-nucleon interaction \cite{beane04}, especially on the
treatment of the scattering lengths larger than the lattice size.
These works examine effects of finite-volume lattice space in
the limit of vanishing lattice spacing, that is, in  the continuum
limit. Here, we focus on effects of the {\it finite lattice
spacing} in the limit of large lattice volume.

Our objective is to determine the parameters of the two-body
interaction on the lattice from known phase shifts. We illustrate
the method in the case of sufficiently low energies, when the
phase shifts can be represented by the ERE parameters. The method
is similar to the continuum case considered in Ref.
\cite{vankolck}, and follows a preliminary attempt involving one
of us (R.S.) several years ago \cite{m-s}. An earlier, related
work can be found in Ref. \cite{dy}. Our results can be applied to
the two-nucleon system at low energies, and the interaction
parameters thus determined are to be used for the many-body Monte
Carlo calculation of Ref. \cite{AS}. In principle, the whole
framework could be used at higher energies, densities and
temperatures, once pion exchange is included explicitly.

The cutoff or renormalization scale is kept at finite values, and
so is the corresponding lattice spacing. As the thermal properties
should be examined at the thermal or infinite-volume limit, the
lattice results needed are at the limit of large lattice space.
This is usually achieved by performing many-body Monte Carlo
calculations with various lattice sizes, followed by applying the
method of finite-size scaling \cite{fss}.  Upon application of the
method, there is no need to consider explicit dependence on the
lattice size in our potential parameters. Determination of the
parameters is greatly simplified at the limit of large space size.
In fact, we find that the basic algebra is the same as that in
free space, apart from the use of the reaction (K) matrix instead
of the standard scattering (T) matrix.

The paper is organized as follows. We briefly discuss the K matrix
as a description of the two-body interaction in a closed space in
Sect. \ref{secK}. In Sect. \ref{secRel}, the relation between the
effective range parameters and the potential parameters is
obtained by the use of a diagrammatic expansion of the K matrix.
An alternative derivation by the direct use of the wave function
is given in App. \ref{AppA}. The case of a large, discrete lattice
is treated in Sect. \ref{secDiscrete}. An elaboration of the
mathematical treatment of the Green's function in this case is
given in App. \ref{AppB}. A brief discussion of our results in
comparison with L\"{u}scher's method and some other concluding
remarks are presented in Sect. \ref{secConcl}.

\section{K (Reaction) matrix in closed space}
\label{secK}

Our method is based on essentially the same scattering formalism
as the well-known L\"{u}scher method \cite{luescher} is. We find
that the use of the reaction, or K, matrix (also termed the
reactance matrix, or R, matrix) \cite{newton,gw}, whose language
is more familiar to the nuclear physics community, greatly
simplifies the formalism.

We consider two particles of mass $M$
interacting through a potential $V({\Vr})$.
The wave function for the relative motion, $\psi_{\Vp}({\Vr})$,
satisfies the Schr\"{o}dinger equation,
\begin{equation}
-(\nabla^2/M) \psi_{\Vp}({\Vr}) +V(\Vr) \psi_{\Vp}({\Vr})
= E_{p} \psi_{\Vp}({\Vr})\;,
\label{schroedinger}
\end{equation}
with $E_{p} ={\Vp}^2/M = p^2/M$ ($p \equiv |{\Vp}|$).
As Eq.~(\ref{schroedinger}) is of second order,
$\psi_{\Vp}({\Vr})$ can be set to describe the
physical state of interest as a combination of two independent
solutions by appropriately choosing boundary conditions. The
standard choice of boundary condition is
that the wave function has, apart from the incident plane wave,
 an outgoing wave
with the asymptotic form $\exp(ipr)/r$, or (though less popular)
an incoming wave with $\exp(-ipr)/r$, either one providing the T
matrix, the usual scattering amplitude. Another choice is for the
wave function to have a standing-wave form,
a combination of the two asymptotic forms. More explicitly, the
wave function in the $\ell$-th angular momentum state has the
asymptotic form:
\begin{equation}
R_{\ell}(pr) \rightarrow j_{\ell}(pr)
-\frac{1}{2}\left(S_{\ell}(p)-1\right)h^{(1)}_\ell(pr)\nonumber
\end{equation}
for the T matrix, or
\begin{equation}
R_{\ell}(pr) \rightarrow j_{\ell}(pr) -p K_{\ell}(p)
n_\ell(pr)\nonumber
\end{equation}
for the K matrix.  Here, $K_{\ell}(p) \equiv
-(i/p)(S_{\ell}(p)-1)/(S_{\ell}(p)+1)$ is the K matrix, and
$S_{\ell}(p)=e^{2i\delta_\ell}$ is the S matrix expressed in terms
of the corresponding phase shift $\delta_\ell$.
$h^{(1)}_\ell(pr)=j_{\ell}(pr)+i n_{\ell}(pr)$ is the spherical
Bessel function of the third kind \cite{abm}. Note that the first
term of the spherical Bessel function $j_{\ell}(pr)$ in the above
equations forms the incident plane wave. From $R_{\ell}(pr)$, the
total wave functions is constructed as
\begin{equation}
\psi_{\Vp}({\Vr})=\sum_{\ell=0}^{\infty}(2\ell+1)i^\ell
R_{\ell}(pr) P_\ell(\cos\theta)\;, \label{schbc}
\end{equation}
where $P_\ell(\cos\theta)$ is the Legendre polynomial with
$\theta$ the angle between ${\Vp}$ and ${\Vr}$.

Clearly, the choices of the outgoing and incoming boundary
conditions are unsuited for the description of two particles
interacting in a closed space.  The choice of the standing-wave
boundary condition can be made by requiring  $V(\Vr)$
and $\psi_{\Vp}({\Vr})$ to satisfy
periodic conditions, such as those that make a cubic box
of size $L\times L \times L$
into a torus,
\begin{eqnarray}
V({\Vr}+{\Vn}L) &=& V({\Vr})\;,
\nonumber\\
\psi_{\Vp}({\Vr}+{\Vn}L)&=&\psi_{\Vp}({\Vr})\;,
\label{wfbc}
\end{eqnarray}
where
$\Vn$ is an integer vector with its components covering a
set of all integer values.
Equation (\ref{wfbc}) restricts
the allowed values of $\{\Vp\}$ to be discrete. In fact,
the Green's
function satisfying Eq.~(\ref{wfbc}) with the standing-wave
boundary condition is written as
\begin{equation}
 G_{\rm P}(\Vp,{\Vr}-{\Vr'}) \equiv \frac{1}{L^3}\sum_{{\Vp'}(\neq{\Vp})}
\frac{\phi_{\Vp'}({\Vr})\phi^*_{\Vp'}({\Vr'})}{E_{p}-E_{p'}}\;,
\label{gsw}
\end{equation}
and obeys
\begin{equation}
[-\nabla^2/M-{\Vp}^2/M] G_{\rm P}(\Vp,{\Vr}-{\Vr'})
=-\frac{1}{L^3}\sum_{{\Vp'}(\neq{\Vp})}
\phi_{\Vp'}({\Vr})\phi^*_{\Vp'}({\Vr'})
=-\delta(\Vr'-\Vr)+\phi_{\Vp}({\Vr})\phi^*_{\Vp}({\Vr'})\;,
\end{equation}
where $\phi_{\Vp}({\Vr}) \equiv {\exp}(i{\Vp}\cdot{\Vr})$.
Here, $\Vp'=2\pi\Vn/L$ is the undisturbed (by $V(\Vr)$) momentum,
while $\{\Vp\}$ forms a discrete set of eigenmomenta in the
closed space, which are determined through a decomposition of the
above Green's function as elaborated in Ref. \cite{luescher}.

In this work, we examine the large-lattice limit by letting
$L\rightarrow \infty$.  In this limit, the periodic condition of
Eq.~(\ref{wfbc}) becomes ineffective and imposes no special
condition, $\Vp$ becoming a continuous spectrum bounded by
the inverse of the finite lattice spacing. The $\Vp'$ sum becomes an integral,
\begin{equation}
\frac{1}{L^3}\sum_{{\Vp'}(\neq{\Vp})} \rightarrow
 \wp\int\frac{d^3 p'}{(2\pi)^3}\;,
\label{linf}
\end{equation}
where $\wp$ stands for the principal value of the integral,
excluding the contribution from $\Vp' = \Vp$.  The range of the
integration in Eq.~(\ref{linf}) is also restricted by the inverse of
the finite lattice spacing.

Our method of using the K matrix is equally applicable to both
a large, closed space and (open) free space.  In fact the
formalism and basic algebra are the same. The K matrix
$K({\Vp'},{\Vp})$ is defined \cite{gw} in terms of
$\psi_{\Vp}({\Vr})$ with the boundary condition (\ref{schbc})
as
\begin{equation}
K({\Vp'},{\Vp}) \equiv \int d^3 r \phi^*_{\Vp'}({\Vr}) V(\Vr)
\psi_{\Vp}({\Vr})\;,
\label{defK}
\end{equation}
and it satisfies the integral equation
\begin{equation}
K({\Vp'},{\Vp}) = V({\Vp'},{\Vp})+ \wp\int\frac{d^3 p''}{(2\pi)^3}
V({\Vp'},{\bm p''}) \bar{G}({\bm p''};{\Vp}) K({\bm p''},{\Vp})\;.
\label{intK}
\end{equation}
Here, $V({\Vp'},{\Vp})$ and $\bar{G}({\Vp'},{\Vp})$ are related to
$V(\Vr)$ and $G_{\rm P}(\Vp,{\Vr}-{\Vr'})$ as
\begin{eqnarray}
V({\Vp'},{\Vp}) &=& \int d^3 r {\phi^*_{\Vp'}({\Vr})V(\Vr)
\phi_{\Vp}({\Vr})} \;, \label{v_mon}\\
(2\pi)^3 \delta^3(\Vp''-\Vp') \wp \bar{G}({\Vp'};{\Vp})&=& \int
d^3 r \int d^3 r' {\phi^*_{\Vp''}({\Vr})G_{\rm
P}(\Vp,{\Vr}-{\Vr'}) \phi_{\Vp'}({\Vr'})}\nonumber\\
&=& (2\pi)^3 \delta^3(\Vp''-\Vp') \wp\frac{M}{p^2-{p'}^2}\;,
\label{g_mon}
\end{eqnarray}
respectively. Note $\wp$ implies $\Vp \neq \Vp'$ in this case.

Equation (\ref{intK}) is the same integral equation that the
standard T matrix $T({\Vp'},{\Vp})$ satisfies for scattering in
free space, except for the Green's function satisfying the
standing-wave boundary condition.  Because the two equations are
of the same structure, the diagrammatic expansions generated from
them, as expansions in terms of $V(\Vp',\Vp)$, are also of the
same structure, apart from the presence of the $+ip$ term
appearing in the T matrix. This term comes from the $\Vp'=\Vp$
contribution that is included in the T-matrix Green's function
(usually denoted as $G^{(+)}({\bm p'},{\Vp})$ for the outgoing
boundary condition). The $+ip$ term is vital for the T matrix to
satisfy the unitarity condition, while the term is not present in
the K matrix, as the K matrix is Hermitian.

Successive substitution of Eq.~(\ref{v_mon}) into Eq.~(\ref{intK})
yields a diagrammatic expansion of the on-shell K matrix. For
getting the expansion, however, we must regulate Green's function
and related (momentum-space) integrals, as discussed in Sect.
\ref{secRel}.

$K({\Vp'},{\Vp})$ is expanded in angular momentum states,
\begin{equation}
K({\Vp'},{\Vp}) = -\frac{4\pi}{M} \sum_{\ell} (2\ell+1)
P_\ell(\hat{\Vp}\,'\cdot\hat{\Vp}) K_\ell(p',p)\;, \label{Kexp}
\end{equation}
where $\hat{\Vp}\,'$ and $\hat{\Vp}$ are unit momentum vectors.
The coefficient $(-4\pi/M)$ is introduced so that the on-shell
$K_\ell(p,p)$ is expressed in terms of the $\ell$-th phase shift
$\delta_\ell$ as
\begin{equation}
K_\ell(p,p)\equiv K_\ell(p)= \frac{1}{p}\tan\delta_\ell(p)\;.
\label{keldef}
\end{equation}
$K_\ell(p)$ is a real function of $p^2$ that is known to be analytic
around $p=0$, so it can be written as the effective range expansion
with a convergence radius of $p^2 \approx 1/(2R)^2$ in the case of
scattering from a potential of the range $R$ \cite{newton}. The
S-wave expansion relevant to this work is
\begin{equation}
K_0^{-1}(p) = p \cot \delta_0(p) = -\frac{1}{a_0} + \frac{1}{2}r_0
p^2 +\mathcal{O}(p^4)\;, \label{effexp}
\end{equation}
where $a_0$ and $r_0$ are the s-wave scattering length and the
effective range, respectively.

\section{Relation between effective-range and potential
parameters using K matrix}
\label{secRel}

In this section, we express the effective-range parameters in
terms of the potential parameters using the K matrix without
specifying the regularization method.
As noted in Sect. \ref{secK},
our method of the K matrix is equally applicable to a large,
closed space and to (open) free space.  The algebra is the
same except for the details associated with regularization.  Without
specifying the regularization method, we can then compare our
method to the previous works based on diagrammatic expansions of
the T matrix, which use different regularization methods
\cite{vankolck,ksw,gegelia}.  In order to solidify the comparison, in
App. \ref{AppA} we also show the derivation of the same results using
the wave function, instead of the diagrammatic expansion, starting
from the definition of the K matrix, Eq.~(\ref{defK}). The
explicit case of the lattice regularization (for a large, closed
space) will be discussed in Sect. \ref{secDiscrete}.

We consider the case where the two particles
interact through a
short-range potential, which is expressed in the form of effective
field theory, consisting of a combination of $\delta^3(\Vr)$ and
powers of the nucleon momentum (square) ${\Vp}^2$,
\begin{equation}
V({\Vr}) = c_0(\Lambda) \delta^3({\Vr})-c_2(\Lambda) [
\nabla^2\delta^3({\Vr})+ \delta^3 ({\Vr})\nabla^2] + \ldots \;,
\label{mompot}
\end{equation}
where the parameters $c_0$, $c_2$, {\it etc.} depend on the cutoff
scale, $\Lambda$. (In the case of a periodic condition
(\ref{wfbc}), $\delta^3(\Vr)$ in $V(\Vr)$ of Eq.~(\ref{mompot}) is
to be replaced by a sum over $\Vn$ of $\delta^3({\Vr}+{\Vn}L)$.)
In momentum space,
\begin{equation}
V({\Vp'},{\Vp}) =  c_0+c_2(p^2+{p'}^2)+ \ldots
\label{v_mon-con}
\end{equation}

Here we show explicitly only the leading terms in the potential
for the case of interest: low-energy phenomena
dominated by the S-wave interaction.
We do not show explicitly higher-order terms
such as the P-wave term
$\loarrow{\nabla}\cdot\roarrow{\nabla}$
or relativistic corrections proportional to ${\Vp^4}$.
A more complete discussion of the various terms
can be found in Ref. \cite{vankolck}.
The potential (\ref{mompot}) is generated
by removing from the theory  other degrees of freedom,
whose effects are now subsumed in $c_0$,
$c_2$, and higher-order counterterms.
For example, in the case the particles are nucleons,
pion-interaction effects can be effectively included in
contact interactions
for $|\Vp| < m_{\pi}/2$, where $m_{\pi}$ is the pion mass.

The potential (\ref{v_mon-con}) is singular,
in the sense that it requires that the problem be regulated.
For example, an integral of the Green's function of
Eq.~(\ref{g_mon}) becomes
\begin{equation}
\wp\int\frac{d^3{\bm p'}}{(2\pi)^3}\bar{G}({\bm p}';{\bm p})
\rightarrow M\wp\int\frac{d^3{\bm p'}}{(2\pi)^3} \frac{F({\bm
p'}^2/\Lambda^2)}{p^2-{\bm p'}^2} \equiv I_0(p,\Lambda) \;
\label{i00}
\end{equation}
by the use of a multiplicative regulator $F(x^2)$, which satisfies
$\lim_{x\to \infty}F(x^2) =0$ and $\lim_{x\to 0}F(x^2) =1$.

For the sake of comparison with other regularization methods, let
us take $F(x^2)$ to be simply an integrable function of $x^2$. We
then have
\begin{eqnarray}
I_0(p,\Lambda)
       &=&-\frac{M}{2\pi^2}
          \left[\int_0^\infty dp' F({p'}^2/\Lambda^2)
  -p^2\cdot\wp\int_0^\infty dp'\frac{F(p'^2/\Lambda^2)}{p^2-p'^2}
          \right] \nonumber\\
       &\equiv&-\frac{M}{2\pi^2} \left[L_1(\Lambda)
       +\frac{p^2}{\Lambda}R((p/\Lambda)^2)\right]\;.
\label{i0}
\end{eqnarray}
Here, $L_1(\Lambda)$ is
\begin{equation}
L_1(\Lambda) \equiv \int_0^\infty  dp' F(p'^2/\Lambda^2)\equiv
\theta_1\Lambda \;,
\label{l1}
\end{equation}
with
\begin{equation}
\theta_1 \equiv \int_0^\infty dx F(x^2)\;, \label{ell1}
\end{equation}
and $R(x^2)$,
\begin{equation}
R(x^2) =\wp\int_0^\infty dx'\frac{F({x'}^2)}{{x'}^2-x^2}\;.
\label{r-func}
\end{equation}
We expect $\theta_1 = \mathcal{O}(1)$, but the exact value depends
on the regulator. $R(x^2)$ is also a regulator-dependent function.
For a sharp cutoff regulator $F(x^2)=\theta(1-x)\theta(x)$, we
have
\begin{equation}
\theta_1 = 1 \;\;{\rm and}\;\; R(x^2)= \frac{1}{2x}\ell
n((1-x)/(1+x)) = 1+\frac{1}{3}x^2+\ldots  \nonumber
\end{equation}

Another regulated integral that appears is
\begin{eqnarray}
I_{2}(p,\Lambda) &=& \wp\int\frac{d^3 {\bm p'}}{(2\pi)^3}
\frac{{\bm p'}^{2}F({\bm p}'^2/\Lambda^2)}{E_{\bm p}-E_{\bm p'}} \nonumber \\
&=&-\frac{M}{2\pi^2}L_{3}(\Lambda) +p^2I_0(p,\Lambda)\;,
\label{i22}
\end{eqnarray}
with
\begin{equation}
L_3 (\Lambda) \equiv 2\pi^2\int \frac{d^3{\bm p'}}{(2\pi)^3}
F({\bm p}'^2/\Lambda^2)
\equiv \theta_3 \Lambda^3
\;. \label{el3}
\end{equation}
For the sharp cutoff regulator,
\begin{equation}
\theta_3 = \frac{1}{3}\;.
\end{equation}

We can define analogous integrals $I_{2n}$,
which satisfy recurrence relations,
\begin{eqnarray}
I_{2n}(p,\Lambda) &\equiv& \wp\int\frac{d^3 p'}{(2\pi)^3}
\frac{p'^{2n}F(p'^2/\Lambda^2)}{E_{\bm p}-E_{\bm p'}} \nonumber \\
&=&-\frac{M}{2\pi^2}L_{2n+1}(\Lambda) +p^2I_{2n-2}(p,\Lambda)\;,
\end{eqnarray}
with
\begin{equation}
L_{2n+1} (\Lambda) \equiv 2\pi^2\int \frac{d^3p'}{(2\pi)^3}
{p'}^{2n-2} F(p'^2/\Lambda^2)\;,
\end{equation}
and thus
\begin{equation}
I_{2n} =
-\frac{M}{2\pi^2}\left[\sum_{i=0}^{n}p^{2i}L_{2(n-i)+1}(\Lambda)+
\frac{p^{2(n+1)}}{\Lambda}R(p^2/\Lambda^2)\right]\;.
\end{equation}

\begin{figure}[tb]
\begin{center}
\includegraphics[width=100mm]{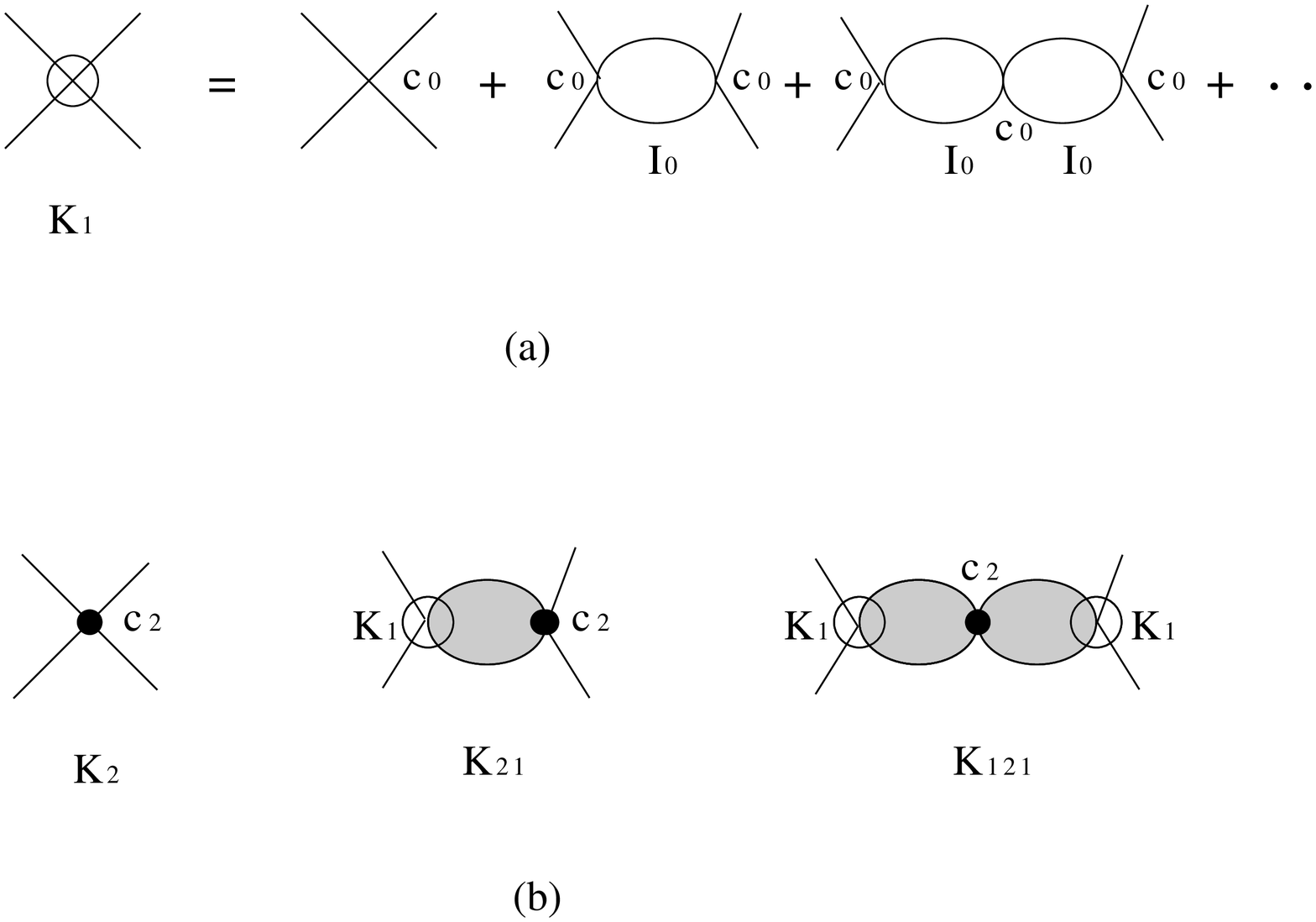}
\caption{Diagrammatic representation of $K_1, K_2, K_{21}$, and
$K_{121}$. The upper figure (a) depicts $K_1$ as a sum of the
iteration of $c_0$, Eq.~(\ref{K1}), with a crossing point as $c_0$.
The open bubble is the regulated Green's function $I_0$.
The lower figure (b) shows $K_2, K_{21}$, and $K_{121}$ as in
Eq.~(\ref{Kps}), with an open circled vertex as $K_1$ and with a
dot vertex as $c_2$.  The shaded bubble is a regulated Green's
function weighted with vertex momenta, $I_2$.} \label{Fig:diagram}
\end{center}
\end{figure}

The diagrammatic expansion of $K({\Vp},{\Vp})$ is depicted in
Fig. \ref{Fig:diagram}.
In the case of large S-wave scattering length $a_0$,
$c_0(\Lambda)I_0(p,\Lambda)$ is close to unity
and diagrams of
all orders in $c_0(\Lambda)$ must be included.
We denote by
$K_1$ the sum of the $c_0(\Lambda)$ contributions,
\begin{equation}
K_1 \equiv c_0(\Lambda)+ c_0(\Lambda) I_0(p,\Lambda)c_0(\Lambda)+
c_0(\Lambda)I_0(p,\Lambda)c_0(\Lambda)I_0(p,\Lambda)c_0(\Lambda)
+\ldots \label{K1}
\end{equation}
On the other hand,
$c_2(\Lambda)$ should be treated perturbatively \cite{vankolck}.
We denote the sets of diagrams with one insertion of $c_2(\Lambda)$ by
$K_2, K_{21},$ and $K_{121}$
if they have, respectively, no $c_0(\Lambda)$ factors,
$c_0(\Lambda)$ factors either before or after the  $c_2(\Lambda)$ insertion,
and $c_0(\Lambda)$ factors both before and after the  $c_2(\Lambda)$ insertion.
The procedure can easily be extended to higher orders.
We have
\begin{eqnarray}
-\frac{4\pi}{M} K_0(p)
 &=& K_1 + K_2 + 2 \cdot K_{21} + K_{121} +
       \mathcal{O}(c_2^2,p^4)\;.
        \label{K0diag}
\end{eqnarray}

After some algebra, we find (suppressing the explicit showing of
the $p$ and $\Lambda$ dependence for a while)
\begin{eqnarray}
K_2 &=& 2 p^2 c_2 \;,\nonumber\\
K_{21} &=& K_1 \cdot c_2(I_2+p^2I_0)\;,\nonumber\\
K_{121} &=& 2 K_1^2 c_2 I_0 I_2\;. \label{Kps}
\end{eqnarray}
Because
\begin{equation}
2 \cdot K_{21} + K_{121} = 2 K_1 \frac{c_2}{c_0}(I_2 K_1 + c_0 p^2
I_0)\;, \nonumber
\end{equation}
we obtain for $K_0(p)$,
\begin{eqnarray}
-\frac{M}{4\pi}K_0^{-1}(p) &=& K_1^{-1}\frac{1}{1 + 2
(c_2/c_0)(p^2+ K_1 I_2)}
+\mathcal{O}(c_2^2,p^4) \nonumber\\
                &=&
\frac{1}{c_0}+\frac{M}{2\pi^2}\left[L_1+\frac{p^2}{\Lambda} R((p/\Lambda)^2)
\right]
-2 \frac{c_2}{c_0} \left[\frac{p^2}{c_0}-\frac{M}{2\pi^2}L_3\right]
+ \mathcal{O}(c_2^2,p^4)
\nonumber\\
                &=&
\frac{1}{c_0}+\frac{M}{2\pi^2}\left(L_1+2\frac{c_2}{c_0}L_3\right)
+\left(\frac{M}{2\pi^2\Lambda}R(0) -2\frac{c_2}{c_0^2}\right)p^2
 + \mathcal{O}(c_2^2, p^4)\;.
\label{K0exp}
\end{eqnarray}
We emphasize that in this derivation, $1-c_0I_0$, {\it not} $c_0$,
is treated perturbatively. We add here a note that Eq.
(\ref{K0exp}) can be also written as
\begin{equation}
 -\frac{4\pi}{M}K_0(p) = \frac{c_0 + 2 c_2 p^2}{1 - (c_0 I_0 + 2 c_2 I_2)} +
       \mathcal{O}(c_2^2,p^4)\;,
\label{nice}
\end{equation}
with the understanding that $1-c_0I_0$ (but {\it not} $c_0$) and
$c_2$ are treated perturbatively.  Up to the $\mathcal{O}(c_2)$
order, Eq.~(\ref{nice}) has the same structure as that of $K_0(p)$
obtained \cite{ksw} by the power divergence subtraction (PDS)
scheme (with their definition of $c_2$ being half of ours).

The S-wave scattering length and effective range are thus
expressed as
\begin{eqnarray}
\frac{M}{4\pi}\frac{1}{a_0} &=& \frac{1}{c_0(\Lambda)} +
\frac{M}{2\pi^2}\left(L_1(\Lambda)
+2\frac{c_2(\Lambda)}{c_0(\Lambda)}L_3(\Lambda) \right)+\ldots
\nonumber\\
\frac{M}{16\pi} r_0 &=& \frac{c_2(\Lambda)}{c_0^2(\Lambda)} -
\frac{M}{4\pi^2}\frac{1}{\Lambda}R(0) +\ldots \;, \label{effrange}
\end{eqnarray}
where terms up to next-to-leading order are explicitly shown.
Equation~(\ref{effrange}) is in agreement with
Ref.~\cite{vankolck}.  Note that terms beyond this
order must include potential terms of $p^4$ and higher.

In this case of a large S-wave scattering length,
the dimensionless parameter $c_0(\Lambda)\theta_1 \Lambda
M$ is near its (unstable) fixed point $-2\pi^2$ \cite{weinberg},
\begin{equation}
c_0(\Lambda) =
\frac{4\pi}{M}\left[\frac{1}{a_0}-\frac{2\theta_1\Lambda}{\pi}\right]^{-1}
+\ldots \approx -\frac{2\pi^2}{\theta_1 M \Lambda}\;,
\end{equation}
while it flows to the trivial
fixed point at the zero value for
$\pi/(2\theta_1 a_0) \gg \Lambda \rightarrow 0$ \cite{birse}.
This observation is consistent with the counting rule that we have
followed: Since our non-relativistic Hamiltonian is a momentum
expansion based on power-counting rules \cite{vankolck}, it is an
expansion about the trivial point.  As the leading term of the
expansion about $p^2 = 0$, the delta-function potential has to be
treated nonperturbatively in order to describe the physics near
the unstable fixed point, away from the trivial one.

\section{Closed, large, discrete lattice}
\label{secDiscrete}

We now examine the effective range expansion on a closed, large,
{\it discrete} lattice. We consider a spatial lattice that is
simply cubic with lattice spacing $a$ and volume $L^3 = (aN)^3$,
in the limit of $N \rightarrow \infty$.

The coordinate is discretized in units of $a$,
\begin{equation}
\Vr \rightarrow a{\bm n}\;, \label{disr}
\end{equation}
where ${\bm n}=\sum_i n_i \hat{\bm r}_i$ is again a vector with
integer components $n_i$ along the directions given by the
Cartesian unit vectors $\hat{\bm r}_i$, $i=1, 2, 3$. Since
\begin{equation}
\int d^3r \rightarrow \sum_{\bm n} a^3 \;,
\end{equation}
we have also
\begin{equation}
\delta^3(\Vr)  \rightarrow \frac{1}{a^3}\delta_{{\bm n},0}\;.
\end{equation}

The range of momenta is limited to the first
Brillouin zone,
\begin{equation}
-\frac{\pi}{a} \leq p_i \leq  \frac{\pi}{a} \;,
\label{brill}
\end{equation}
for each momentum component $i$. In the limit $N\to \infty$ the
momentum is continuum in this interval. The wave functions in
coordinate and momentum spaces are related by
\begin{equation}
\psi(a{\bm n})=
\int^{\pi/a}_{-\pi/a}\int^{\pi/a}_{-\pi/a}\int^{\pi/a}_{-\pi/a}
\frac{dp_1dp_2dp_3}{(2\pi)^3} \tilde{\psi}(\Vp) e^{ia {\bm
n}\cdot\Vp} \nonumber\;,
\end{equation}
which is seen to satisfy the required periodicity even for a
finite $\Vn$, Eq.~(\ref{wfbc}) with Eq.~(\ref{disr}).

When the standard four-point difference formula is used, the
kinetic-energy operator is expressed on the cubic spatial lattice
as
\begin{eqnarray}
 -\nabla^2 \psi_{\Vp}({\Vr})&\rightarrow&
-\sum_{i=1}^3\frac{1}{a^2}\left\{
             \psi\left((\bm{n}+\hat{\bm r}_i)a\right)
            +\psi\left((\bm{n}-\hat{\bm r}_i)a\right)
            -2\psi(a \bm{n})\right\}\nonumber\\
&=&
\int^{\pi/a}_{-\pi/a}\int^{\pi/a}_{-\pi/a}\int^{\pi/a}_{-\pi/a}
\frac{dp_1dp_2dp_3}{(2\pi)^3}\tilde{\psi}(\Vp)e^{i {\bm n}\cdot
\Vp }\frac{1}{a^2}P(\Vp)\;.
\end{eqnarray}
That is, as an operator in the spatial space, ${\Vp}^2$ is
represented as
\begin{equation}
{\Vp}^2 \rightarrow \frac{1}{a^2}P(\Vp) \equiv
\frac{2}{a^2}\sum_{i=1}^3(1-\cos(a p_i))\;. \label{p2op}
\end{equation}
$P(\Vp)/a^2$ becomes ${\Vp}^2$ in the continuum limit $a\rightarrow 0$,
while for a finite $a$, we have
\begin{equation}
          0 \leq \frac{1}{a^2}P(\Vp) \leq \frac{12}{a^2} \;.
\end{equation}
$P(\Vp)=0$ occurs only when $\Vp=0$, and thus we have no problem
of fermion doubling. Note that Eq.~(\ref{p2op}) leads to the
formal operator expression for the regulator,
\begin{equation}
 F \left(\frac{a^2 p'^2}{\pi^2}\right)
 = \left(\frac{a}{\pi}\right)^2
 \int^{\pi/a}_{-\pi/a}\int^{\pi/a}_{-\pi/a}\int^{\pi/a}_{-\pi/a}
\frac{dp''_1dp''_2dp''_3}{(2\pi)^3} \;
 \delta\left(\left(\frac{ap'}{\pi}\right)^2
 - \frac{2}{\pi^2}\sum_{i=1}^3
 \left[1-\cos\left(a p''_i\right)\right]\right)
\;,
 \label{latreg}
\end{equation}
showing explicitly that a cubic spatial symmetry is now imposed.
This suggests the identification
\begin{equation}
\frac{\pi}{a} \sim \Lambda \;. \label{momcutoff}
\end{equation}

We then write
\begin{eqnarray}
I_0(p,a)&\equiv&-\frac{M}{2a}\frac{1}{(2\pi)^3}
\wp\int^{\pi}_{-\pi}\int^{\pi}_{-\pi}\int^{\pi}_{-\pi}
\frac{dxdydz}{3 - a^2p^2/2 - (\cos x - \cos y - \cos z)}\nonumber \\
&\equiv&-\frac{M}{2\pi^2}\left[L_1\left(\frac{\pi}{a}\right) +
\frac{ap^2}{\pi}R\left(\left(\frac{pa}{\pi}\right)^2\right)\right]\;.
\label{Ninfty}
\end{eqnarray}
Here,
\begin{equation}
L_1\left(\frac{\pi}{a}\right) = \frac{1}{8\pi a}
\wp\int^{\pi}_{-\pi}\int^{\pi}_{-\pi}\int^{\pi}_{-\pi}
\frac{dxdydz}{3-(\cos x +\cos y +\cos z)} \equiv
\frac{\pi}{a}\theta_1\;, \label{L1lat}
\end{equation}
where $\theta_1$ is introduced analogously to Eq.~(\ref{l1}), and
\begin{equation}
R\left(\left(\frac{pa}{\pi}\right)^2\right)=\frac{1}{16}
\wp\int^{\pi}_{-\pi}\int^{\pi}_{-\pi}\int^{\pi}_{-\pi}
\frac{dxdydz}{[3-(\cos x +\cos y +\cos z)]
              [3-a^2p^2/2-(\cos x +\cos y +\cos z)]} \;.
\label{Rlat}
\end{equation}

The evaluation of these integrals requires some care. $I_0(p,a)$
of Eq.~(\ref{Ninfty}) is related to Watson's triple integral
$\mathcal{I}_3^p(z)$, discussed in App. \ref{AppB}, through
\begin{equation}
I_0(p,a)= -\frac{M}{2a}\; \mathcal{I}_3^p\left(3-\frac{a^2p^2}{2}\right) \;.
\end{equation}
Watson's integral $\mathcal{I}_3^p(z)$ has branch points at $z=\pm3$,
so the limit of small
$ap/\pi$ is delicate.
In App. \ref{AppB} we find
\begin{equation}
\mathcal{I}_3^p\left(3-\epsilon\right)
= A + B\epsilon
  + \mathcal{O}\left(\epsilon^2\right)\;,
\end{equation}
with $A=0.505462\cdots$ and $B=0.0486566\cdots$.
Therefore, $\theta_1$ is
\begin{equation}
\theta_1 = \pi A = 1.58796\cdots\;,
\label{theta1}
\end{equation}
in agreement with Ref. \cite{dy} (where $\theta_1\equiv 2\pi/\eta$
with $\eta = 3.956\cdots$). In addition, the $R$ function is
\begin{equation}
R\left(x^2\right) =\frac{{\pi}^3}{2} B + \mathcal{O}\left(x^2\right)
= 0.754330 \cdots
+ \mathcal{O}\left(x^2\right)\;.
\label{Rterm}
\end{equation}

Higher-order integrals can be obtained as in open space.
For example,
\begin{eqnarray}
I_{2}(p,a) &\equiv& -\frac{M}{a^3}\frac{1}{(2\pi)^3}
\wp \int^{\pi}_{-\pi}\int^{\pi}_{-\pi}\int^{\pi}_{-\pi}dxdydz
\frac{3 - (\cos x+\cos y+\cos z)}{3 - a^2p^2/2 - (\cos x+\cos y+\cos z)}
\nonumber \\
&\equiv&-\frac{M}{2\pi^2}L_3\left(\frac{\pi}{a}\right)
+p^2I_0(p,a)\;.
\label{moreNinfty}
\end{eqnarray}
Here,
\begin{equation}
L_3\left(\frac{\pi}{a}\right) = \frac{1}{4\pi a^3}
\int^{\pi}_{-\pi}\int^{\pi}_{-\pi}\int^{\pi}_{-\pi}dxdydz
=\frac{2\pi^2}{a^3}
\equiv \theta_3\left({\frac{\pi}{a}}\right)^3\;,
\label{L3lat}
\end{equation}
so
\begin{equation}
\theta_3=\frac{2}{\pi}\;.
\label{theta3}
\end{equation}
For comparison, as noted in Sect. \ref{secRel}, a sharp
momentum-cutoff regulator gives $\theta_1=1,\; R(0)=1$, and
$\theta_3=1/3$.

Following the same algebra as described in the previous section,
we then find the inverse of the S-wave K matrix $K_0^{-1}(p)$
expressed in terms of the potential parameters $c_0(a)$ and
$c_2(a)$, combined with Eq.~(\ref{effexp}), as
\begin{eqnarray}
  K_0^{-1}(p)&=& -\left(\frac{4\pi}{M}\right)\left\{
\frac{1}{c_0(a)}+\frac{M}{2\pi^2}\left[L_1\left(\frac{\pi}{a}\right)
+\frac{ap^2}{\pi} R\left(\left(\frac{pa}{\pi}\right)^2\right)
\right]\right.
\nonumber\\
&& \qquad\qquad
\left. -2 \frac{c_2(a)}{c_0(a)} \left[\frac{p^2}{c_0(a)}
-\frac{M}{2\pi^2}L_3\left(\frac{\pi}{a}\right)\right]
+ \mathcal{O}(c_2^2,p^4)\right\}
\nonumber\\
 &=& -\frac{1}{a_0} + \frac{1}{2}r_0 p^2 +\mathcal{O}(p^4)\;.
\label{K0exp2}
\end{eqnarray}

Let us first examine the case $c_2 = 0$. In terms of $a$ and
$\theta_1$, the K matrix is expressed in the same form as that for
the continuum. The scattering length $a_0$ is given as
\begin{equation}
\frac{1}{a_0} = \frac{4\pi}{M c_0(a)} + \frac{2\theta_1}{a} \;
\label{1a0lat}
\end{equation}
in the $\mathcal{O}((a p/\pi)^0)$ order in the power counting. The
$c_0(a)$ in this lowest order, $c_0^{(0)}(a)$, is then
\begin{equation}
c_0^{(0)}(a) = \frac {4\pi}{M}
\left(\frac{1}{a_0}-\frac{2}{a}\theta_1\right)^{-1} \approx
-\frac{2\pi^2}{\theta_1 M}\left(\frac{a}{\pi}\right)\;.
\label{c00lat}
\end{equation}
Here, the approximated expression corresponds to that at the fixed
point, and is valid when
\begin{equation}
\left|\frac{a}{a_0}\right| \ll 2\theta_1 = 3.17591 \cdots \;.
\label{inequal}
\end{equation}

When $c_2$ is included,  with Eqs. (\ref{L1lat}), (\ref{Rlat}),
and (\ref{L3lat}) we obtain
\begin{eqnarray}
\frac{M}{4\pi}\frac{1}{a_0} &=& \frac{1}{c_0(a)} +
\frac{M}{2\pi^2}\left(\theta_1\frac{\pi}{a} + 2\theta_3
\left(\frac{\pi}{a}\right)^3\frac{c_2(a)}{c_0(a)} \right)+\ldots
\;,
\nonumber\\
\frac{M}{16\pi} r_0 &=& \frac{c_2(a)}{c_0^2(a)} -
\frac{Ma}{4\pi^3}R(0)+\ldots
\label{a0r0}
\end{eqnarray}
$c_0(a)$ and $c_2(a)$ are determined from $a$ and $r_0$ by
inverting Eq. (\ref{a0r0}). Up to $\mathcal{O}(c_2)$ order, we
have
\begin{eqnarray}
c_0(a) &=& c_0^{(0)}(a)\left\{1+\frac{M^2r_0 }{16a^3}\theta_3 \, \eta \,
(c_0^{(0)}(a))^2\right\} \;,
\nonumber\\
c_2(a) &=& \frac{Mr_0 }{16\pi}\eta \, (c_0^{(0)}(a))^2 \;,
\label{c02eft}
\end{eqnarray}
where $c_0^{(0)}(a)$ is given by Eq. (\ref{c00lat}), and
\begin{equation}
\eta\equiv 1 + \frac{4a}{\pi^2 r_0}R(0) \approx 1\; ,
\label{eta}
\end{equation}
the last approximation valid when
\begin{equation}
\left|\frac{a}{r_0}\right| \ll \frac{\pi^2}{4 R(0)} = 3.27098
\cdots \;. \label{inequal2}
\end{equation}

We now examine numerically the validity of the expressions for
$c_0(a)$ and $c_2(a)$, Eq. (\ref{c02eft}), in the case the
particles are nucleons. Both spin singlet and triplet scattering
lengths for S-wave two-nucleon scattering
are known to be large \cite{m}: $a_{0s} = -23.740 \pm 0.020$ fm
and $a_{0t} = +5.419 \pm 0.007$ fm, respectively (for the
neutron-proton system). In contrast, the spin singlet and triplet
effective ranges have more natural sizes: $r_{0s} = 2.77 \pm 0.05$
fm and $r_{0t} = 1.753 \pm 0.008$ fm, respectively.
 These values are comparable with the range $1/m_\pi$
of the interaction, which sets the expected limit of validity of
the ERE,
\begin{equation}
  p \leq \frac{1}{|r_0|} = 0.37 - 0.57 \;\;{\rm fm^{-1}}
  \equiv p_{max} \;.
\label{papprox}
\end{equation}
For optimal results, the momentum cutoff $\Lambda \sim \pi/a$
should be set greater than $p_{max}$. This requires
\begin{equation}
  \left|\frac{a}{r_0}\right| \leq \pi \;,
\label{amax}
\end{equation}
so that
the lattice spacing $a$ should be less than about
$\pi/p_{max} = 8.5 - 5.5$ fm,
which is a lax limit.
The inequalities (\ref{amax}), (\ref{inequal}), and
(\ref{inequal2}) can be realized in nuclear systems with a fairly
wide range of
lattice spacings.

Using Eq.~(\ref{c00lat}), we have
\begin{equation}
c_0(a) \approx c_0^{(0)}(a)\left[ 1 + \frac{\pi^2 \theta_3
r_0}{4\theta_1^2 a}\eta\right] \;. \label{c02eftprime}
\end{equation}
The second term in the square brackets
can be large if the lattice spacing is small. As a numerical
example, let us take $a = 2.0$ fm, which corresponds to
$\pi/a =1.57 \; {\rm fm}^{-1}$. We find
$c_0(a)\approx 1.8 \, c_0^{(0)}(a)$, that is, the value of
$c_0(a)$ increases by
about 80\% by the inclusion of the momentum-dependent term in the
potential.  While this is important when going to next order in a
calculation, it does {\it not} imply a failure of the EFT
expansion. The parameters of the EFT Lagrangian, such as $c_0(a)$,
are not directly observable. The convergence of the expansion is
assured as long as Eq. (\ref{amax}) is satisfied. In fact, the
ratio
\begin{equation}
\left|\frac{c_2(a)}{c_0(a)}\right|p_{max}^2
\approx \frac{a}{8\theta_1 r_0}
        \left[ \eta^{-1}+\frac{\pi^2\theta_3r_0}{4\theta_1^2a}\right]^{-1}
\end{equation}
is numerically small: For example, for $a = 2.0$ fm, it amounts to
about 0.06. This ratio in fact vanishes in the continuum limit.
The perturbative treatment of $c_2(a)$ seems to be reasonable and
the effective range expansion is properly described by the
momentum-dependent potential, Eq. (\ref{mompot}).

\section{Discussion and Conclusion}
\label{secConcl}

We have expressed the effective range parameters in terms of the
potential parameters up to ${\Vp}^2$ on a large, discrete
lattice, basically in the same way as in free space.
Equation (\ref{K0exp2}) relates two sets of the parameters,
\begin{equation}
 \{c_0(a)\; {\rm and} \; c_2(a)\}\;\; {\rm and} \;\;
 \{a_0\; {\rm and} \; r_0\}
\label{parameters}
\end{equation}
in the $L\to\infty$ limit.

In the case of a finite $L$, the relation is complicated
because the momentum spectrum $\{\Vp\}$ depends on $L$ and also on
$c_0(a)$ and $c_2(a)$,
\begin{equation}
\{\Vp\} \rightarrow \{{\Vp}\;[L;c_0(a),c_2(a)]\}
\;.\label{monspec}
\end{equation}
As well as depending on $L$, $\{\Vp\}$ is discrete because only
a discrete set of
standing waves satisfying the periodic boundary condition,
Eq.~(\ref{wfbc}), can exist in a finite closed space.  The exact
spectrum of $\{\Vp\}$ must be determined numerically for the given
$c_0(a)$ and $c_2(a)$. This is the basic procedure of
L\"{u}scher's method.  It involves an elaborate computation to
relate $a_0$ and $r_0$
(or phase shifts at $\{\Vp\}$)
directly from the energy (mass) spectrum extracted from lattice
computations (as usually attempted in the case of lattice QCD).
L\"{u}scher's well-known formula to relate $a_0$ and the lowest
energy for a given $L$ is, for example, a perturbative expansion
about $\Vp = 0$ at the limit of $a \rightarrow 0$ \cite{luescher},
and corresponds to the simplest relation coming out of
Eqs.~(\ref{parameters}) and (\ref{monspec}).

Our objective was to obtain the relation (\ref{parameters}). For
this, we took the closed space to be large, by letting $L
\rightarrow \infty$ in the torus space. $\{\Vp\}$ is then
continuous without the limit $a \rightarrow 0$ and, for a finite
$a$, is limited to the first Brillouin zone, Eq.~(\ref{brill}).
Note a subtle, but perhaps basic point in this work, concerning
the $N \rightarrow \infty$ limit that we have taken. For a simple
cubic lattice, rotational invariance is broken by both the
ultraviolet cutoff $a$ and the infrared cutoff $L$. For a large
value of $N$, however, the discrete version of the kinetic
operator $P(\Vp)/a^2$, $P_N(\Vp)/a^2$, becomes
\begin{equation}
\frac{1}{a^2}P_N({\bm k}) \equiv
\frac{2}{a^2}\sum_{i=1}^3(1-\cos(2\pi k_i/N) =
\left(\frac{2\pi}{Na}\right)^2{\Vk}^2 + \mathcal{O}\left(
               \frac{k_i^4}{N^4a^2}\right) \;. \label{discmom3}
\end{equation}
As $N \rightarrow \infty$, the rotational symmetry is approached
in $P_N(\Vp)/a^2 \rightarrow P(\Vp)/a^2$ for $\Vp \rightarrow 0$.
Though we approach the infinitely large closed space by
maintaining the spatial cubic symmetry (by increasing $N$ for each
of the three spatial components), the corresponding momentum
spectrum near $\Vp =0$ effectively approaches that of spherical
symmetry. In other words, by removing the infrared cutoff, only
momenta near the ultraviolet cutoff know of the breaking of
rotational invariance, see Eq. (\ref{latreg}). The regularization
at $N \rightarrow \infty$ meets the requirement of preservation of
the proper symmetry for the momenta of interest in the low-energy
theory. We emphasize, thus, that we take the $L=aN \rightarrow
\infty$ limit, which differs from the $N\rightarrow \infty$ limit
with finite $L$.

We can then expand the K matrix in terms of ${\Vp}^2$,
Eq.~(\ref{K0exp2}), and obtain the relation between the two sets
of parameters in Eq.~(\ref{parameters}) through a direct
comparison of the expansions power by power, without carrying out
L\"{u}scher's elaborate algebra. Here, for the ${\Vp}^2$ expansion
involving $c_0$ and $c_2$, one must be careful so as to meet the
power-counting rules associated with the application of effective
field theory. As described in this way, our method simply amounts
to the standard effective range expansion (with some caveats). Our
method of the K matrix is indeed equally applicable to both of a
large, closed space and (open) free space with the same algebra,
as elaborated in Sect. \ref{secRel} and App. \ref{AppA}.

In conclusion, following the appropriate counting rules for the
S-wave nucleon-nucleon interaction, we obtained Eqs.
(\ref{c00lat}), (\ref{c02eft}), and (\ref{eta}) for a large,
simple cubic lattice, where Eqs. (\ref{theta1}), (\ref{Rterm}),
and (\ref{theta3}) hold. In principle the same method can be
pushed beyond the effective range expansion, through the explicit
inclusion of pions. The expressions so obtained tell us how
low-energy two-nucleon data determine the dependence of EFT
parameters on the lattice spacing, and can be applied to Monte
Carlo calculations of many-nucleon systems in large lattices.

\bigskip
\bigskip
\bigskip

\noindent{\bf Acknowledgments}

We acknowledge stimulating discussions with Boris Gelman, Dean
Lee, and Rob Timmermans at the initial stages of this work. We are
grateful to George Weiss for correspondence regarding Ref.
\cite{MM}. RS thanks Martin Savage for clarifying his work and
some critical issues on the subject, and also Toru Takahashi for
explaining L\"{u}scher's method and its applications to lattice
QCD. UvK thanks the Nuclear Theory Group and the Institute for
Nuclear Theory at the University of Washington, and the Kellogg
Lab at Caltech for hospitality during the time this work was
carried out. This work was supported in part by the U.S.
Department of Energy under grants DE-FG03-87ER40347 (RS) and
DE-FG02-04ER41338 (UvK), by the U.S. National Science Foundation
under grant 0244899
(at Caltech, RS),
and by the Alfred P. Sloan Foundation (UvK).


\appendix

\section{Derivation of Effective range expansion in large space,
using of the wave function}
\label{AppA}

We are going to apply the integral equation,
\begin{equation}
\psi_{\Vp}({\Vr}) = \phi_{\Vp}({\Vr}) + \int d^3 r' G_{\rm
P}({\Vp},{\Vr}-{\Vr'}) V(\Vr') \psi_{\Vp}({\Vr'}) \;,
\label{int-eq}
\end{equation}
with $G_{\rm P}({\Vp}, \Vr-\Vr')$ of Eq.~(\ref{gsw}) with Eq.~(\ref{linf}).
The method followed here is essentially the same as that of Ref. \cite{cohen}.

 In order to clarify the derivation, let us first consider
the potential with $c_2 = 0$:
\begin{equation}
V(\Vr) = c_0(\Lambda) \delta^3 ({\Vr}) \;. \label{simpleV}
\end{equation}
With this potential, Eq.~(\ref{defK}) yields gives
\begin{equation}
 K({\Vp},{\Vp}) = c_0(\Lambda) \psi_{\Vp}(\Vr =0)
 \;.
\label{Kwave1}
\end{equation}
$\psi_{\Vp}(\Vr =0)$ is determined as follows:
Equation~(\ref{int-eq}) gives
\begin{equation}
 \psi_{\Vp}({\Vr}) = e^{i\Vp\cdot\Vr}
 + c_0(\Lambda)M \psi_{\Vp}(\Vr = 0)\cdot\wp\int\frac{d^3p'}{(2\pi)^3}
 \frac{e^{i\Vp'\cdot\Vr}}{p^2-\Vp'^2}\;,
\label{int-eq2}
\end{equation}
or,
\begin{equation}
 \psi_{\Vp}({\Vr}=0) = 1
 + c_0(\Lambda) I_0(p,\Lambda) \psi_{\Vp}({\Vr}=0)\;,
\end{equation}
by the use of the regulated Green's function Eq. (\ref{i00}).
We thus find
\begin{equation}
 \psi_{\Vp}({\Vr}= 0) = \frac{1}{1
- c_0(\Lambda)I_0(p,\Lambda)}\; \label{wf0}\;.
\end{equation}
Equation (\ref{wf0}) confirms that $\psi_{\Vp}({\Vr}=0)$ represents an
S wave. From Eqs. (\ref{Kexp}) and (\ref{Kwave1}) with
Eqs.~(\ref{wf0}), (\ref{i0}), and (\ref{l1}), we obtain
\begin{eqnarray}
K_0(p) &=&-\frac{M}{4\pi}c_0(\Lambda) \psi_{\Vp}(\Vr =0)
\nonumber \\
&=& -\frac{Mc_0(\Lambda)}{4\pi}\left\{ 1 +
\frac{Mc_0(\Lambda)}{2\pi^2}\Lambda\left[\theta_1
+\left(\frac{p}{\Lambda}\right)^2
R\left((p/\Lambda)^2\right)\right]\right\}^{-1} \;. \label{k0p}
\end{eqnarray}
The effective range expansion of Eq. (\ref{effexp}) relates
$c_0(\Lambda)$ to the scattering length $a_0$,
\begin{equation}
\frac{1}{a_0} = \frac{4\pi}{M c_0(\Lambda)} +
\frac{2\theta_1\Lambda}{\pi} \;. \label{1a0}
\end{equation}

We now consider the potential of Eq.~(\ref{mompot}).  Substituting
it into Eq. (\ref{defK}), we obtain
\begin{equation}
 K({\Vp},{\Vp}) =
 [c_0(\Lambda)+c_2(\Lambda) {\Vp}^2]\psi_0({\Vp},\Lambda)
 -c_2(\Lambda)\psi_2({\Vp},\Lambda) \;,
\label{Rwave2}
\end{equation}
where $\psi_0 ({\Vp},\Lambda)$ and $\psi_2 ({\Vp},\Lambda)$
stand for
$\psi_{\Vp}({\Vr} \rightarrow 0)$ and $\nabla^2\psi_{\Vp}({\Vr}
\rightarrow 0)$ with a cutoff $\Lambda$, respectively.
Following
the same procedure as the one for Eq.~(\ref{simpleV}) above, we
find that $\psi_0({\Vp},\Lambda)$ and $\psi_2({\Vp},\Lambda)$
satisfy the coupled linear equations,
\begin{eqnarray}
\left[1-c_0 I_0-c_2 I_2\right] \psi_0 + c_2 I_0 \psi_2 &=& 1  \;, \nonumber \\
\left[c_0 I_2+c_2I_4\right] \psi_0 +
\left[1-c_2I_2\right]\psi_2&=& - p^2 \;. \label{couplewf}
\end{eqnarray}
(For simplicity, we suppress the $\Vp$ and $\Lambda$ dependence in
$\psi$'s, $c$'s, and $I$'s in the rest of this appendix.)
Equation
(\ref{couplewf}) yields
\begin{eqnarray}
\psi_0 &=&\left[(1-c_2I_2)+c_2I_0p^2\right]/Det\nonumber\\
\psi_2 &=&-\left[c_0I_2+c_2I_4 +(1-c_0I_0-c_2I_2)p^2\right]/Det\;,
\end{eqnarray}
where
\begin{equation}
Det=1-c_0I_0-2c_2I_2+c_2^2I_2^2-c_2^2I_0I_4\;.
\end{equation}
$K_0(p)$ is then
\begin{equation}
-(4\pi/M)K_0(p)
=\left[c_0+c_2^2I_4+2(c_2-c_2^2I_2)p^2+c_2^2I_0 p^4 \right]/Det
\;,
\end{equation}
which is exact, obtained from Eq. (\ref{couplewf}).

We now impose power counting rules by treating $c_2$
perturbatively and by expanding about $1-c_0I_0$.
We obtain
\begin{eqnarray}
-(M/4\pi)K_0^{-1}(p) &\approx&
\frac{1}{c_0}\left[1-c_0I_0-2c_2I_2-2\frac{c_2}{c_0}(1-c_0I_0)p^2\right]
+\mathcal{O}(c_2^2,p^4)\nonumber\\
&=&
\frac{1}{c_0}+\frac{M}{2\pi^2}\left(L_1+2\frac{c_2}{c_0}L_3\right)
+\left(\frac{M}{2\pi^2\Lambda}R(0)
-2\frac{c_2}{c_0^2}\right)p^2
+\mathcal{O}(c_2^2,p^4)\;.
\nonumber\\ &&
\end{eqnarray}
We thus recover Eq.~(\ref{effrange}), in agreement with
Ref.~\cite{vankolck}.

\section{Watson's triple integral}
\label{AppB}

We define a function of a complex variable $z$,
$\mathcal{I}_3(z)$, as
\begin{equation}
\mathcal{I}_3(z) =
\frac{1}{(2\pi)^3}\int_{-\pi}^{\pi}\int_{-\pi}^{\pi}\int_{-\pi}^{\pi}
\frac{d^3 \phi}{z-\lambda({\bm \phi})}\;,
\end{equation}
where ${\bm \phi}=(\phi_1,\phi_2,\phi_3)$ and
\begin{equation}
\lambda({\bm \phi})=\cos\phi_1+\cos\phi_2+\cos\phi_3\nonumber\;.
\end{equation}
When $|z| \leq 3$, the integrand has poles at the values of ${\bm
\phi}$ satisfying $\lambda({\bm \phi})=z$.  These poles generate
in $\mathcal{I}_3(z)$
two branch points at $z=\pm 3$ and a branch cut between the two
points.  Because of this structure on the
$z$ complex plane, the (asymptotic) expansions about $z=\pm 3$ are
complicated.  The expansion about $z=3-0$ is what we would like to
find, and for obtaining the K-matrix expansion, we need to
consider the principal value of the integral.  Note that if we
were to naively expand the integrand, we would find that all
coefficients of $(3-z)^n$ ---except for $n=0$--- in
\begin{equation}
\mathcal{I}_3(z) = \sum_{n=1}(3-z)^n\cdot
\frac{1}{(2\pi)^3}\int_{-\pi}^{\pi}\int_{-\pi}^{\pi}\int_{-\pi}^{\pi}
\frac{d^3 \phi}{(3-\lambda({\bm \phi}))^{n+1}}
\end{equation}
diverge with the degree of the divergence worsening as $n$
increases.  A mathematical complication here is that each term in
the expansion of the principal-valued integral has to be evaluated
numerically.

Previously, the triple integral $\mathcal{I}_3(z)$ at $z=3+0$ was
analyzed by Watson \cite{watson}, and for $|z| \geq 3$ was
studied \cite{MM,MW,watson-note,ID} in connection to random walks on lattices
and to lattice dynamics in condensed matter.  The expansion about
$z = 3+0$ was found to be \cite{MM,MW}
\begin{equation}
\mathcal{I}_3(z) = A -\frac{1}{\sqrt{2}\pi}(z-3)^{1/2}
-B(z-3) +\frac{1}{4\sqrt{2}\pi}(z-3)^{3/2}+\ldots \;,
\label{i3exp}
\end{equation}
where $A$, $B$, {\it etc.} are the coefficients of the integer
powers. We have found that these coefficients have been quoted
sometimes incorrectly in the literature: The first term
$\mathcal{I}_3(3)=A$ has been expressed analytically in terms of
Gamma functions \cite{watson-note,ID} but in apparent disagreement
with the correct numerical value \cite{MM,ID}
\begin{equation}
A=0.505462\cdots \; .
\label{A}
\end{equation}
We also find
the coefficient of the third term
to be
\begin{equation}
B=0.0486566\cdots \; ,
\label{B}
\end{equation}
instead of the value $0.014625\cdots$ quoted in Ref. \cite{MM}.
In the following and the rest of this paper, we use the values
that we believe to be correct.

Analytic continuation of $\mathcal{I}_3(z)$, Eq.~(\ref{i3exp}),
from the region $|z| \geq 3$ to the region $|z|< 3$, above and
below the branch cut, yields $\mathcal{I}_3^+(z)$ and
$\mathcal{I}_3^-(z)$, respectively:
\begin{equation}
\mathcal{I}_3^{\pm}(z) = A \pm \frac{i}{\sqrt{2}\pi}(3-z)^{1/2}
+B(3-z) \mp \frac{i}{4\sqrt{2}\pi}(3-z)^{3/2}+\ldots
\end{equation}
The Plemelj formula \cite{carrier} then gives
\begin{eqnarray}
\mathcal{I}_3^p(z)&\equiv&
\frac{1}{(2\pi)^3}\wp\int_{-\pi}^{\pi}\int_{-\pi}^{\pi}\int_{-\pi}^{\pi}
\frac{d^3 \phi}{z-\lambda({\bm \phi})} \nonumber\\
&=&\frac{1}{2}(\mathcal{I}_3^+(z) + \mathcal{I}_3^-(z)) = A +
B(3-z) +\mathcal{O}((3-z)^2) \label{final}
\end{eqnarray}
near $z=3$ for $z \leq 3$.

In the following, we sketch the derivation of Eq.~(\ref{i3exp})
because the literature describing the derivation \cite{MM} is
difficult to locate, and also because our value of the coefficient
of the third term disagrees with the original one quoted in
Ref. \cite{MM}, as noted above.  We first write
\begin{equation}
 \mathcal{I}_3(3+\epsilon) = \int_0^{\infty} dt
 e^{-(3+\epsilon)t}I_0^3(t)\;,
\end{equation}
where
\begin{equation}
I_0(t) \equiv \frac{1}{\pi}\int_0^{\pi}dx e^{(\cos x)t}
\label{i0div}
\end{equation}
is the modified Bessel function \cite{abm}.
The $t$ integral in $\mathcal{I}_3(z)$ can be divided into two integrations,
$[0,T]$ and $[T,\infty]$ for a large numerical value of $T$:
\begin{equation}
 \mathcal{I}_3(z)= \mathcal{I}_3^{a}(z) + \mathcal{I}_3^{b}(z)\;.
\end{equation}
$\mathcal{I}_3^{a}(z)$ is expanded about $z=3$,
\begin{equation}
\mathcal{I}_3^{a}(3+\epsilon)=\int_0^{T} dt
 e^{-3t}I_0^3(t)-\epsilon \int_0^{T} dt\;
 t e^{-3t}I_0^3(t)+\ldots \;,
\label{ia}
\end{equation}
and is numerically computed for each term in the $\epsilon$
expansion using the closed form of $I_0(t)$,
Eq.~(\ref{i0div}).
In $\mathcal{I}_3^{b}(3+\epsilon)$ we use instead the asymptotic expansion
\begin{equation}
I_0(t) =\frac{e^t}{\sqrt{2\pi t}}
        \left(1+\frac{1}{8t}+\frac{9}{128t^2}+\ldots\right)\;,
\label{i0divas}
\end{equation}
or,
\begin{equation}
I_0^3(t) = \frac{e^{3 t} }{(2\pi t)^{3/2}} \sum_{i=0} \frac{d_i}{t^i}\;,
\label{i03divas}
\end{equation}
with
\begin{equation}
d_0=1, \; d_1=\frac{3}{8}, \; d_2=\frac{33}{128}, \; d_3=\frac{281}{1024},
\; \cdots\nonumber\;.
\end{equation}
Introducing the Incomplete Gamma function
\begin{equation}
\phi_m(x) \equiv \int_{1}^{\infty} dt \; t^m e^{-xt}\nonumber
\;,
\end{equation}
which satisfies
\begin{equation}
\phi_{m-1}(x) = \frac{x}{m}\phi_m(x)
-\frac{1}{m}e^{-x}\nonumber
\;,
\quad
\phi_{-1/2}(x) = \frac{2}{\sqrt{\pi}} \int_{\sqrt{x}}^{\infty} du \;
e^{-u^2}\nonumber
\;,
\end{equation}
$\mathcal{I}_3^{b}(3+\epsilon)$
can then be written as
\begin{eqnarray}
\mathcal{I}_3^{b}(3+\epsilon)
&=&\frac{1}{(2\pi)^{3/2}T^{1/2}}\sum_{i=0}\frac{d_i}{T^i}
\phi_{-i-3/2}(\epsilon T)\nonumber \\
&=&\frac{2}{(2\pi)^{3/2}}\left\{
\frac{1}{T^{1/2}}
\sum_{i=0} \frac{d_i}{(2i+1)T^i}
-\sqrt{\pi}d_0{\epsilon}^{1/2}\right.
\nonumber\\
&&\qquad\qquad
\left.+T^{1/2}\left[d_0
-\sum_{i=1} \frac{d_i}{(2i-1)T^i}
\right]\epsilon
+\frac{2}{3}\sqrt{\pi}d_1{\epsilon}^{3/2}
+\ldots\right\} \;.
\label{mess}
\end{eqnarray}
The coefficients of ${\epsilon}^0$ and $\epsilon^1$ are combined
from $\mathcal{I}_3^{a}(3+\epsilon)$ of Eq.~(\ref{ia})
and $\mathcal{I}_3^{b}(3+\epsilon)$ of Eq.~(\ref{mess}), and
are then numerically computed for various large values of $T$. By
examining the numerical results, we obtain the asymptotic expansion of
Eq.~(\ref{i3exp}), with coefficients (\ref{A}) and (\ref{B}).
Note that the terms of half-integer $\epsilon$ powers come only from
$\mathcal{I}_3^{b}(3+\epsilon)$.

\end{document}